\documentclass[galaxies,article,accept,moreauthors,pdftex,10pt,a4paper]{Definitions/mdpi}

\firstpage{1}
\pubvolume{xx}
\issuenum{1}
\articlenumber{5}
\pubyear{2019}
\copyrightyear{2019}
\history{Received: date; Accepted: date; Published: date}

\usepackage{subcaption}

\newcommand{\suml}{\sum\limits}
\newcommand{\fermi}{\textit{Fermi}-LAT}
\newcommand{\Swift}{\textit{Swift}}
\newcommand{\g}{\ensuremath{\gamma}}
\newcommand{\zred}{z_{\rm red}}

\newcommand{\E}[1]{\times 10^{#1}}

 \usepackage[labelformat=simple]{subcaption}

\DeclareCaptionLabelFormat{subcaptionlabel}{\normalfont(\textbf{#2}\normalfont)}
\Title{The VHE $\gamma$-Ray View of the FSRQ PKS~1510-089}

\Author{Michael Zacharias %
 $^{1,2,*}$\orcidA{}, Dijana Dominis Prester $^{3}$, Felix Jankowsky $^{4}$, Elina Lindfors $^{5}$, Manuel~Meyer $^{6}$\orcidE{}, Mahmoud Mohamed $^{4}$, Heike Prokoph $^{7}$, David Sanchez $^{8}$, Julian Sitarek $^{9}$\orcidI, Tomislav Terzic $^{3}$, Stefan Wagner $^{4}$, Alicja Wierzcholska $^{10}$\orcidL, for the H.E.S.S. and~MAGIC~Collaborations}

\AuthorNames{Michael Zacharias, et al.
}

\address{%
$^{1}$ \quad Ruhr Astroparticle and Plasma Physics Center (RAPP Center), Insitut f\"ur theoretische Physik IV, Ruhr-Universit\"at Bochum, D-44780 Bochum, Germany \\
$^{2}$ \quad Centre for Space Science, North-West University, Potchefstroom, 2520 South Africa \\
$^{3}$ \quad Croatian MAGIC Consortium, Rudjer Boskovic Institute, University of Rijeka, University of Split and University of Zagreb, 10000, Zagreb, Croatia; dijana@phy.uniri.hr (D.D.P.); tterzic@phy.uniri.hr (T.T.) \\ 
$^{4}$ \quad Landessternwarte, Universit\"at Heidelberg, K\"onigstuhl, D-69117 Heidelberg, Germany; fjankows@lsw.uni-heidelberg.de (F.J.); mmh@lsw.uni-heidelberg.de (M.M.); swagner@lsw.uni-heidelberg.de~(S.W.)  \\
$^{5}$ \quad Finnish MAGIC Consortium, Tuorla Observatory, University of Turku and Astronomy Division, University~of~Oulu, 90014 Oulun, Finland; elilin@utu.fi \\
$^{6}$ \quad W. W. Hansen Experimental Physics Laboratory, Kavli Institute for Particle Astrophysics and Cosmology, Department of Physics and SLAC National Accelerator Laboratory, Stanford University, \mbox{Stanford, CA 94305, USA}; mameyer@stanford.edu \\
$^{7}$ \quad DESY, D-15738 Zeuthen, Germany; heike.prokoph@desy.de \\
$^{8}$ \quad Laboratoire d'Annecy-le-Vieux de Physique des Particules, Universite Savoie Mont-Blanc, CNRS/IN2P3, 74941 Annecy-le-Vieux, France; david.sanchez@lapp.in2p3.fr \\
$^{9}$ \quad Department of Astrophysics, University of Lodz, 90236 Lodz, Poland; jsitarek@uni.lodz.pl \\
$^{10}$\quad Institute of Nuclear Physics, Polish Academy of Sciences, PL-31342 Krakow, Poland; alicja.wierzcholska@ifj.edu.pl}

\corres{Correspondence: mz@tp4.rub.de}

\abstract{The flat spectrum radio quasar PKS~1510-089 is a monitored target in many wavelength bands due to its high variability. It was detected as a very-high-energy (VHE) $\gamma$-ray emitter with H.E.S.S. in 2009, and has since been a regular target of VHE observations by the imaging Cherenkov observatories H.E.S.S. and MAGIC. In this paper, we summarize the current state of results focusing on the monitoring effort with H.E.S.S. and the discovery of a particularly strong VHE flare in 2016~with H.E.S.S. and MAGIC. While the source has now been established as a weak, but regular emitter at VHE, no correlation with other energy bands has been established. This is underlined by the 2016~VHE flare, where the detected optical and high-energy $\gamma$-ray counterparts evolve differently than the VHE flux.
}


\begin{document}


\section{Introduction}
The correlations between blazar emissions in different energy bands are best probed with long-term monitoring, providing unbiased sampling.
Especially for ground-based observatories this is hard to achieve for even a small number of sources. The {\it Fermi}
 satellite has transformed the monitoring of blazars in the high-energy (HE) $\gamma$-ray band ($E>100\,$MeV) through its continuous surveillance of the whole sky every three hours (although somewhat less uniform after its hardware failure in March 2018) as detailed in \cite{t18,l18}. In the optical 
and radio 
bands many monitoring programs are run thanks to the large number of available telescopes. However, in other energy bands the monitoring capabilities are limited. In the X-ray band the Neil Gehrels \emph{Swift} observatory runs a limited monitoring effort and can follow up on flares. MAXI on board the International Space Station provides all-sky capabilities within 1 orbit with limited sensitivity.
In the very-high-energy (VHE) $\gamma$-ray band ($E>100\,$GeV) the monitoring effort is limited by sensitivity, e.g., for FACT and HAWC \cite{d18}, or by time constraints due to competition with other objects. The latter strongly influences the monitoring efforts of the three large imaging atmospheric Cherenkov telescope (IACT) facilities H.E.S.S., MAGIC and VERITAS. Nonetheless, they have been running limited monitoring projects on a number of blazars.

Here, we report on the ongoing monitoring efforts by H.E.S.S. and MAGIC of the flat spectrum radio quasar (FSRQ) PKS~1510-089. It is located at a redshift $\zred=0.361$ and possesses a bright broad-line region (BLR), e.g., \cite{hjea09,tea12}. Hence, VHE photons produced within the boundaries of the BLR should be absorbed. As the emission region of \g-rays was thought to be close to the central black hole, VHE emission from FSRQs was considered unlikely by many. However, several detections of FSRQs~\cite{magic08,magic11,hess13,magic15,magic16,hess17,magic17} challenge this picture and suggest that jets are able to produce \g-rays also further downstream in the jet.

To verify that these are not simply one-time-only flaring events, but that FSRQs produce VHE emission on all time scales, monitoring programs have been initiated with H.E.S.S. and MAGIC on PKS~1510-089. While these are not unbiased, they have already provided important information. During a strong multiwavelength flaring event in 2015, variability on night-by-night scales at VHE \g-rays was observed for the first time from this source \cite{aMea17,zea16,zea17}. Furthermore, MAGIC observations integrated during low-states in the HE band revealed a significant VHE signal with an average, integrated flux $\bar{F}(E>150\,\mbox{GeV})=(4.3\pm0.6)\E{-12}\,$cm$^{-2}$s$^{-1}$ \cite{acc18}. Hence, PKS~1510-089 is not only variable in VHE \g-rays but also a persistent source. This has a direct and very important consequence: the absorption of VHE photons through the BLR cannot be too severe, and the emission region must be at the edge or even outside of the BLR at all times. This, in turn, implies that the usual model for FSRQ \g-ray emission, namely inverse-Compton scattering of BLR photons, might not be correct.

This paper gives the status of the H.E.S.S. monitoring efforts on PKS~1510-089, and its early results. One of the important outcomes is the detection of an unprecedented VHE flare in 2016, which was also followed-up with MAGIC. Details of this flare are reported here. Additional multiwavelength data are gathered for comparison from \fermi\ in the HE \g-ray band, from \Swift-XRT in the X-ray band and from ATOM \cite{hauserea05} in the R-band.

%
\section{Monitoring with H.E.S.S.} \label{sec:mon}

\begin{figure}[H]
\centering
\centering
\includegraphics[width=0.80\textwidth]{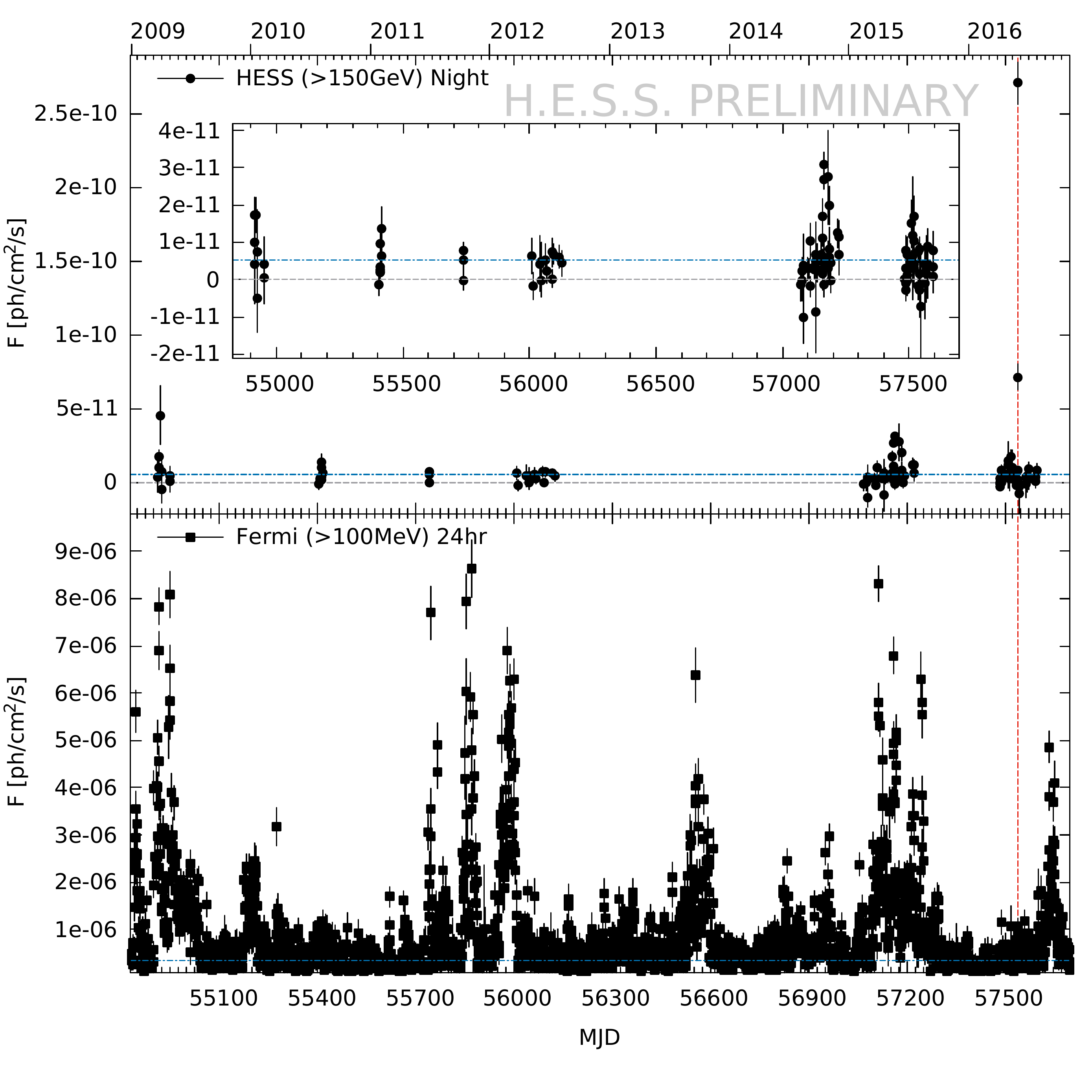}
\caption{Nightwise binned fluxes from 2009 to 2016 in the VHE band taken with H.E.S.S. (top panel) and the HE band taken with \fermi\ (bottom panel). The inset shows a zoom in on VHE fluxes without the 2016 flare. The blue dash-dotted line marks the average flux, while the gray dashed line marks the zero-flux level. The vertical red dashed line marks the time of the VHE flare in 2016.}
\label{fig:mon-all}
\end{figure}

\begin{figure}[H]
\centering
\includegraphics[width=0.80\textwidth]{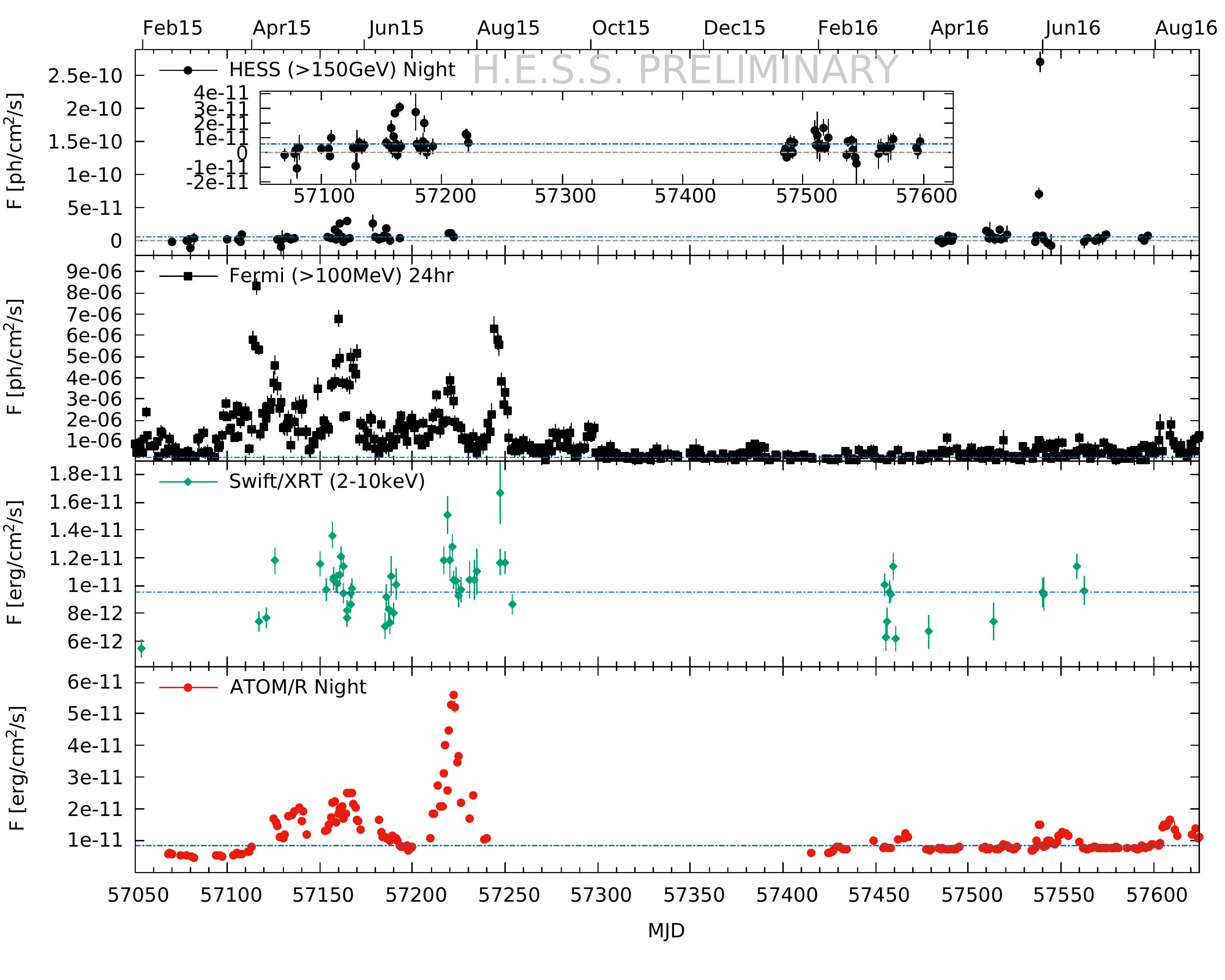}
\caption{Nightwise binned fluxes from 2015 to 2016 in the VHE band taken with H.E.S.S. (1st~panel), the HE band taken with \fermi\ (2nd panel), the X-ray band taken with \Swift-XRT (3rd panel), and~the optical R-band (4th panel). The inset shows a zoom in on VHE fluxes without the 2016 flare. The~blue dash-dotted line marks the average flux, while the gray dashed line marks the zero-flux level.}
\label{fig:mon-1516}
\end{figure}

After the detection in 2009 \cite{hess13}, H.E.S.S. has continued observing PKS~1510-089 with low cadence. Since 2015 this effort has been significantly increased with several hours of observations each month during the visibility period (which typically lasts from February to July each year) resulting in observations almost every night without moon interference. The resulting nightwise lightcurve including all observations is shown in Figure~\ref{fig:mon-all}, and a focus on the 2015 and 2016 season is shown in Figure~\ref{fig:mon-1516}. Note that nightwise bins do not guarantee a significant flux per night due to the limited sensitivity and the dimness of the source in the low state. In fact, about 50\% of the nights shown in Figure~\ref{fig:mon-all} are compatible with zero. The bright VHE flare in 2016 clearly stands out with peak fluxes up to 10 times higher than the previous record holder in 2015. In order to reveal details of the other times, the~inset shows the zoom in on the fluxes without the 2016 flare. The~flare is further discussed in Section~\ref{sec:flare}. The average, integrated flux for the whole time frame is $\bar{F}(E>150\,\mbox{GeV})=(5.1\pm0.3)\E{-12}\,$cm$^{-2}$s$^{-1}$, which is compatible within errors with the MAGIC low-flux level, but includes the bright states, as~well. The average of the 2015--2016 time frame is compatible with the average of the whole time frame. In~both cases, a constant flux is ruled out with very high significance. This is underlined by the fractional variability \cite{eea02}

\begin{align}
 F_{\rm var} = \frac{\sqrt{S^2-\sigma_{\rm err}^2}}{\bar{F}} \label{eq:Fvar},
\end{align}
where $S^2$ is the variance, $\sigma_{\rm err}^2$ is the mean square error, and $\bar{F}$ is the average flux of the source in the considered data set. For the whole data set $F_{\rm var}^{\rm VHE} = 3.2\pm 0.1$, and for the 2015--2016 time frame $F_{\rm var}^{\rm VHE} = 3.3\pm 0.1$. These large values are driven by the 2016 flare. Removing the two nights of that event give $F_{\rm var}^{\rm VHE} = 0.8\pm 0.2$, which still implies significant variability. Defining the variability time scale between two subsequent flux points as \cite{ww95}
\begin{align}
 t_{\rm var} = \bar{F} \frac{t_{i+1}-t_{i}}{|F_{i+1}-F_{i}|} \label{eq:tvar}
\end{align}
the minimum variability time scale is $t_{\rm var}^{\rm VHE} = (0.8\pm0.06)\,$h, which was exhibited during the major flare in 2016. The error on the variability time scale has been derived through error propagation.


The bottom panel in Figure~\ref{fig:mon-all} shows the nightwise HE \g-ray fluxes obtained with \fermi. The~average, integrated flux in this band is $\bar{F}(E>100\,\mbox{MeV})=(3.39\pm0.03)\E{-7}\,$cm$^{-2}$s$^{-1}$ for the whole time frame, and $\bar{F}(E>100\,\mbox{MeV})=(2.58\pm0.05)\E{-7}\,$cm$^{-2}$s$^{-1}$ for 2015-2016. The source has been very active in this band for large parts of the considered time frame with $F_{\rm var}^{\rm HE}~=~1.170\pm 0.005$~for the whole time frame, and $F_{\rm var}^{\rm HE} = 1.11\pm 0.01$ for 2015--2016. The minimum variability time scale is $t_{\rm var}^{\rm HE} = 0.69\pm0.06\,$h for the whole time frame exhibited during a flare in 2011, and $t_{\rm var}^{\rm HE} = 1.3\pm0.2\,$h for the 2015--2016 time frame exhibited in August 2015.

\begin{figure}[H]
\centering
\begin{subfigure}[t]{0.45\textwidth}
\centering
\includegraphics[width=0.9\textwidth]{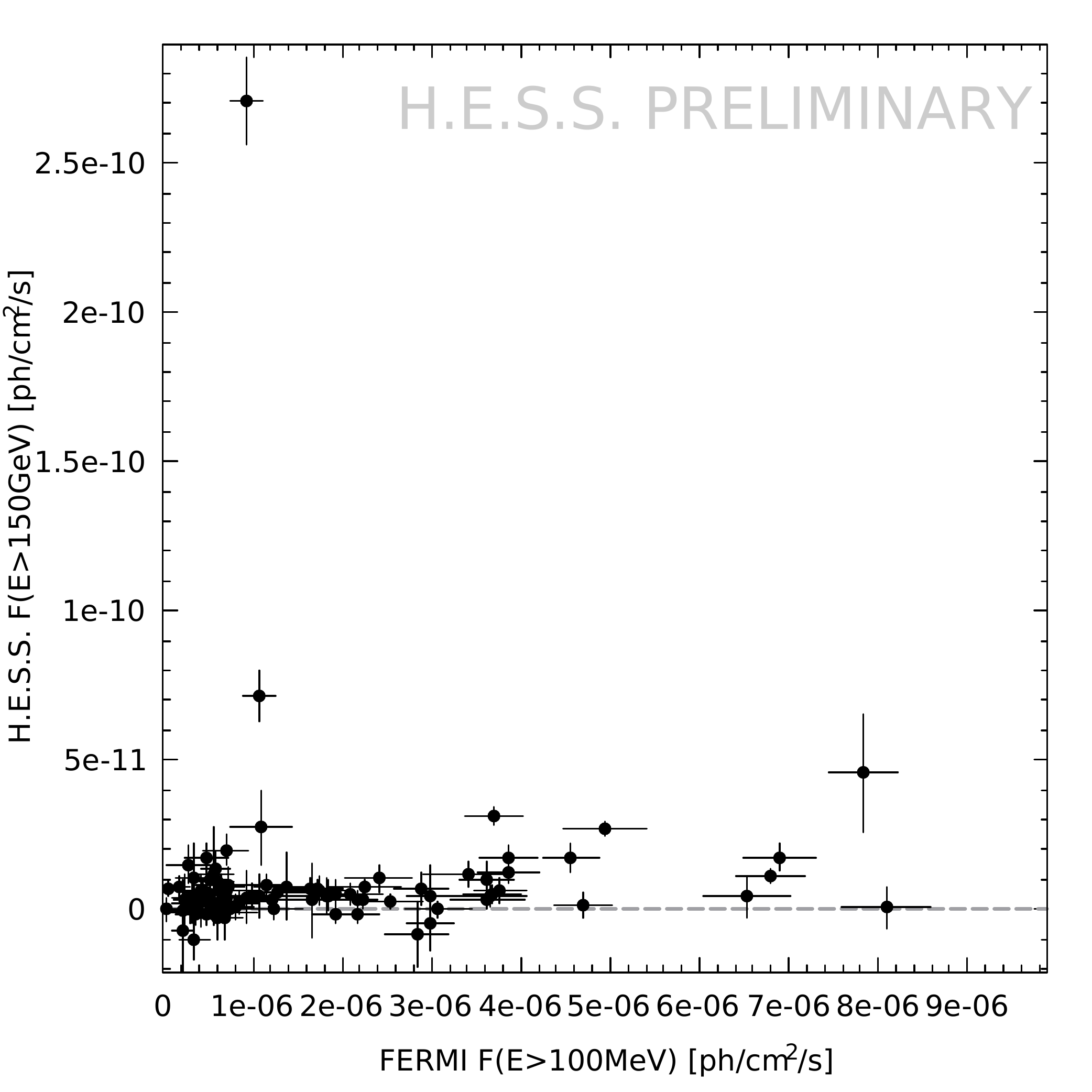}
\caption{Scatterplot showing H.E.S.S. fluxes versus \fermi\ fluxes. The black dashed line marks the zero-flux level for the VHE band.}
\label{fig:scatHF}
\end{subfigure}
\hspace{5mm}
\begin{subfigure}[t]{0.45\textwidth}
\centering
\includegraphics[width=0.9\textwidth]{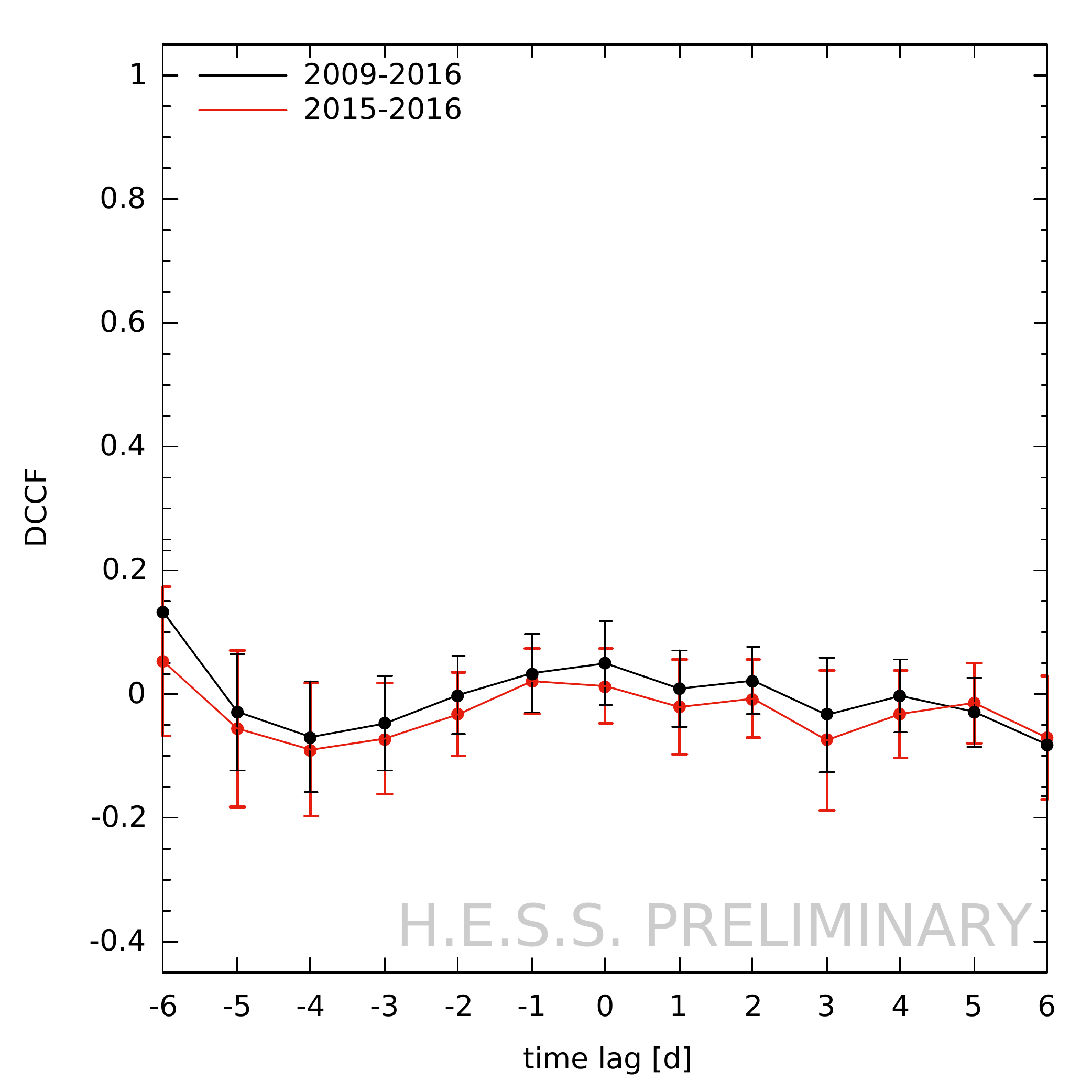}
\caption{DCCF between H.E.S.S. and \fermi\ fluxes for the entire time range (black) and 2015--2016 (red).}
\label{fig:dccfHF}
\end{subfigure}
\vspace{-12pt}
\caption{Data points for simultaneous data from H.E.S.S. and \fermi.}\label{figure3}
\end{figure}

Unfortunately, many of the HE flares were not followed up with with H.E.S.S. due to observational constraints. Nonetheless, it is interesting to investigate whether there is any correlation between these two bands. Plotting the simultaneously recorded fluxes of the two bands against each other can reveal direct correlations. The scatterplot is shown for H.E.S.S. and \fermi\ fluxes in Figure~\ref{figure3}a. The discrete cross-correlation function (DCCF) can uncover correlations with time-delays in non-simultaneous and unevenly spaced data \cite{ek88}:
\begin{align}
 DCCF(\tau) = \frac{1}{N}\suml_{i,j} \frac{(F_i^a - \bar{F}^a)\,(F_j^b - \bar{F}^b)}{S^a\,S^b} \label{eq:dccf}
\end{align}
where $F^a$ and $F^b$ are the fluxes of two lightcurves with mean $\bar{F}^a$ and $\bar{F}^b$ and variance $S^a$ and $S^b$, respectively. The sum goes over all $N$ pairs $i,j$ in the time interval $\tau$.
The DCCF between the VHE and HE \g-ray fluxes is shown in Figure~\ref{figure3}b for the full time frame and 2015-2016, respectively.

The scatterplot does not contain any strong evidence for a direct correlation between VHE (here~integrated above 150 GeV) and HE fluxes in the data. While on several occasions a high HE flux is accompanied with a significant VHE flux, this is not a general rule, as also VHE-flux levels compatible with zero are recorded for similar HE fluxes. On the other hand, similar VHE fluxes can occur at different HE-flux levels. The 2016 VHE flare again stands out for the relatively low simultaneous HE fluxes. The non-correlation of VHE and HE fluxes is also underlined by the flat DCCF. It should be noted that the different integration times (a few hours for H.E.S.S. and $24\,$h for \fermi) might~influence the conclusions here given the relatively fast variability found in this~source.

For the 2015-2016 time frame, data from \Swift-XRT and ATOM have been analyzed, giving~the X-ray and R-band lightcurves for these years. The X-ray average, integrated flux is $\bar{F}(2\,\mbox{keV}~<~E~<~10\,\mbox{keV})~=~(9.5\pm0.1)\E{-12}\,$erg\,cm$^{-2}$s$^{-1}$. The flux is incompatible with a constant flux with high significance, and $F_{\rm var}^{X} = 0.19\pm0.01$. The fastest variability is $t_{\rm var}^{X}= 8\pm4\,$h.\footnote{The large error implies that this value is not highly significant. Trials might reduce the significance further. Hence, this time scale should be regarded as a lower limit.} Given the low cadence in these observations as visible in the third panel of Figure~\ref{fig:mon-1516}, it is difficult to distinguish flares from a ground state. In the optical R-band, the average, integrated flux is $\bar{F}(\mbox{R})=(8.605\pm0.005)\E{-12}\,$erg\,cm$^{-2}$s$^{-1}$. The flux is highly variable with $F_{\rm var}^{R} = 0.679\pm0.002$, and~$t_{\rm var}^{R}= 10.3\pm0.5\,$h. The lightcurve, shown in the bottom panel of Figure~\ref{fig:mon-1516}, reveals a highly active state in 2015 and a mostly quiet state in 2016. In April and May 2015 ATOM recorded correlated activity in the optical band with the HE $\gamma$-ray band. The very bright optical flare in July 2015, which was the brightest flux state ever recorded with ATOM in PKS~1510-089, only had a mild counterpart in the HE band. Compared to the other optical flares in 2015, the July outburst was more than 2 times brighter.

\begin{figure}[H]
\centering
\begin{subfigure}[t]{0.45\textwidth}
\centering
\includegraphics[width=0.90\textwidth]{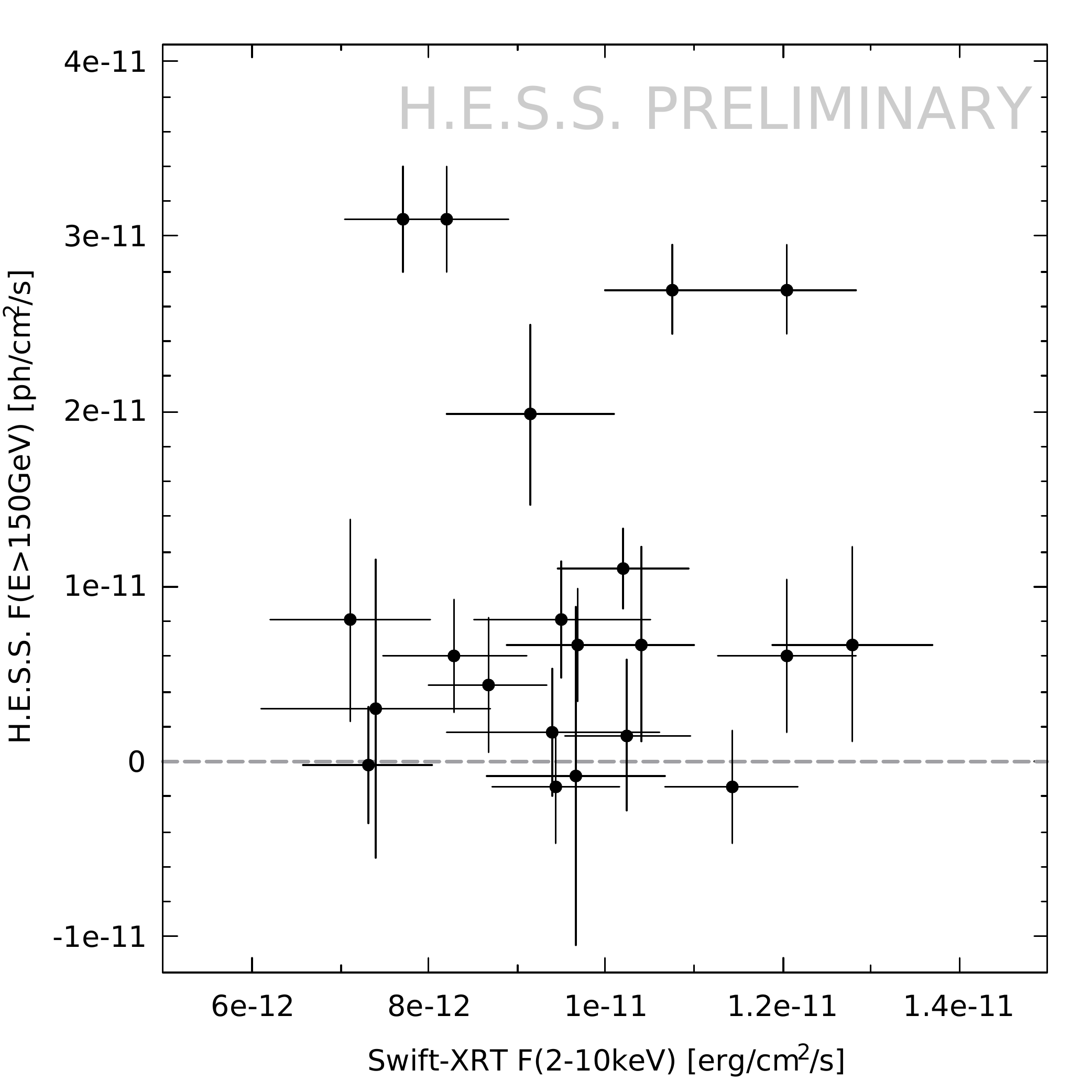}
\caption{Scatterplot showing H.E.S.S. fluxes versus \Swift-XRT fluxes. The black dashed line marks the zero-flux level for the VHE band.}
\label{fig:scatHX}
\end{subfigure}
\hspace{5mm}
\begin{subfigure}[t]{0.45\textwidth}
\centering
\includegraphics[width=0.90\textwidth]{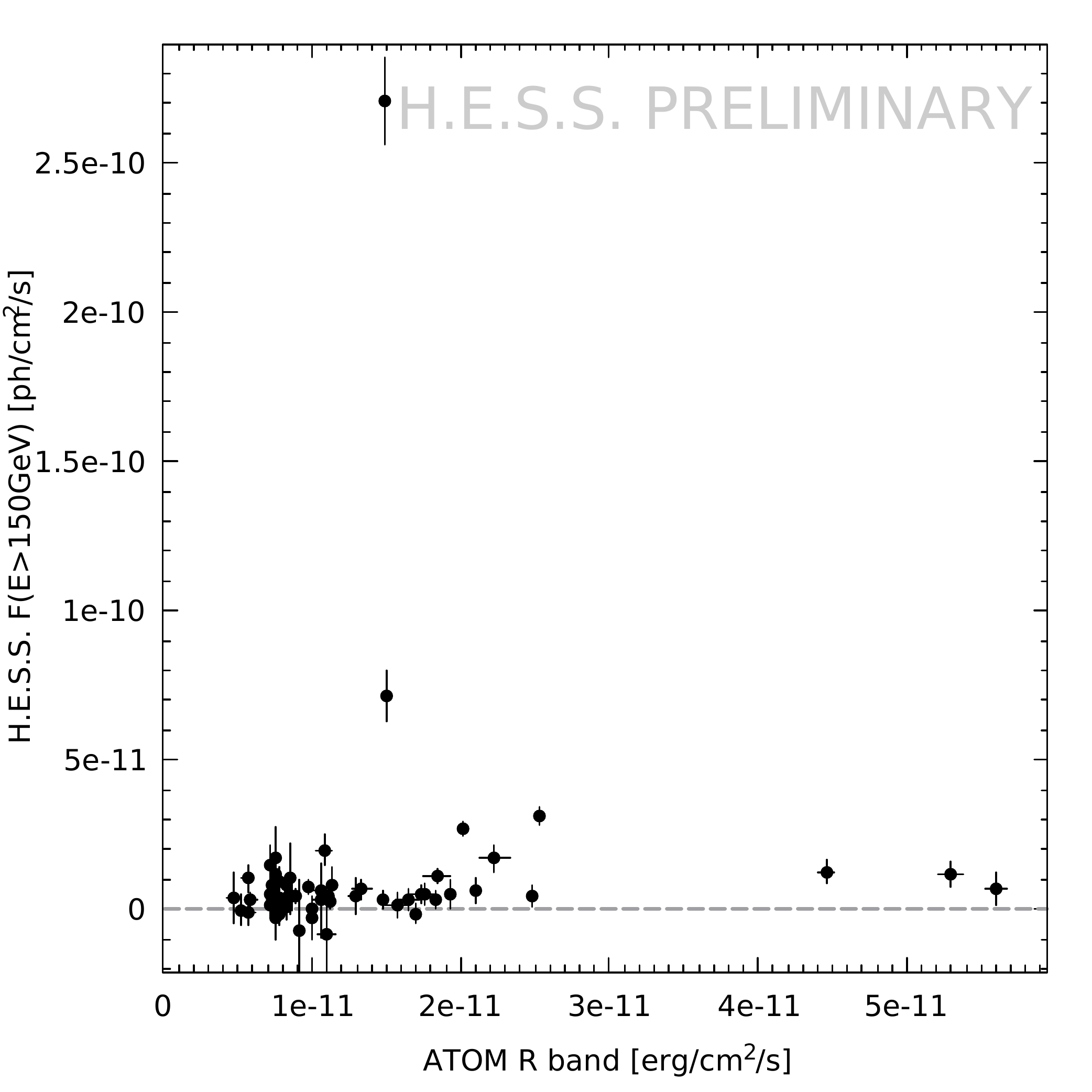}
\caption{Scatterplot showing H.E.S.S. fluxes versus ATOM/R fluxes. The black dashed line marks the zero-flux level for the VHE band.}
\label{fig:scatHA}
\end{subfigure}
\vspace{-12pt}
\caption{Scatterplots for simultaneous data for 2015--2016.}\label{figure4}
\end{figure}

Scatterplots have also been produced for VHE versus X-ray and VHE versus R-band fluxes, which are shown in Figure~\ref{figure4}a,b, respectively. As the number of data points are low, no DCCFs have been calculated. No conclusions can be drawn from the VHE versus X-ray scatterplot at this point. Unfortunately, no X-ray coverage was obtained during the 2016 VHE flare. The VHE versus R-band scatterplot suggests that high optical fluxes (i.e., above $2\E{-11}\,$erg/cm$^2$/s) imply significant VHE fluxes (i.e., fluxes that deviate by more than 1$\sigma$ from zero). The low number of data points makes this a weak conclusion. However, high VHE fluxes do not imply high optical fluxes (using the same threshold), as is demonstrated by the 2016 VHE flare.

%
%
\section{The 2016 VHE Flare} \label{sec:flare}

\begin{figure}[H]
\centering
\centering
\includegraphics[width=0.55\textwidth]{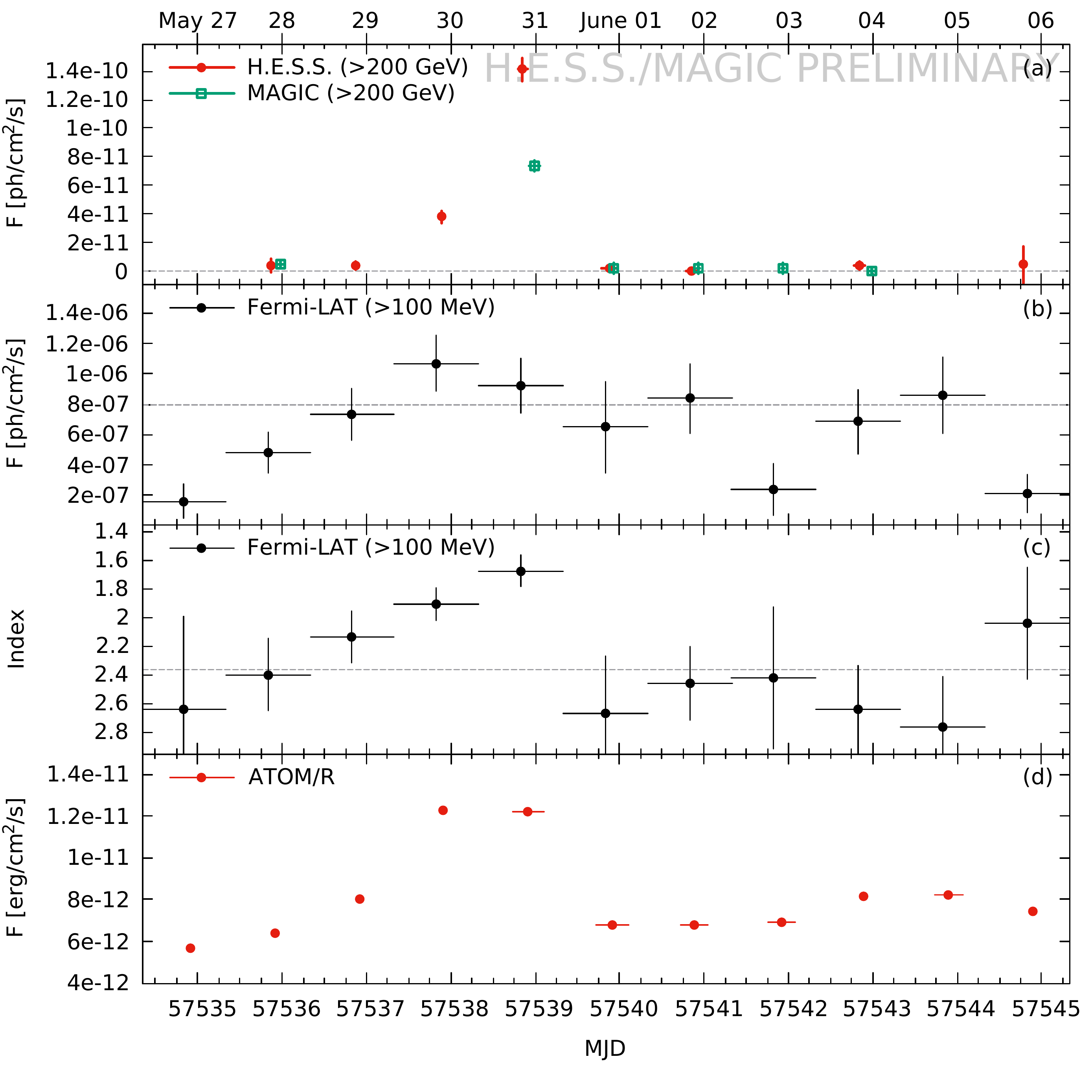}
\caption{(\textbf{a}) VHE lightcurve obtained with H.E.S.S. (red) and MAGIC (green) for $E>200\,$GeV in nightly bins. The dashed line marks the zero flux level. (\textbf{b}) HE lightcurve obtained with \fermi\ for $E>100\,$MeV in daily bins centered on the H.E.S.S. observation window. The dashed line marks the long-term average. (\textbf{c}) HE \g-ray spectral index from \fermi\ observations. The dashed line marks the long-term average. (\textbf{d}) Optical R-band lightcurve obtained with ATOM in nightly bins. In all panels only statistical errors are shown.}
\label{fig:flarelc-all}
\end{figure}

The summary of the monitoring results in Sec.~\ref{sec:mon} has already hinted at the unprecedented nature of the flare in 2016. The nightly lightcurve around this event is shown in Figure~\ref{fig:flarelc-all}. The flare lasted less than 3 days in the VHE band with a peak in the late hours of MJD~57538 (30 May 2016 ---hereafter ``maximum night''). In the HE band a flux rise seems to have happened. However, this is barely significant, as it is hovering around the long-term average and more than a factor 10 below previous flares. On the other hand, the spectral index clearly reduces compared to the average $\sim$2.4, peaking at $\sim 1.6$. Hence, while the integrated flux in the HE band barely changed, the spectrum itself significantly hardened. The optical flux rises by a factor of 2 from the beginning of the event to its peak. However, this again is a much smaller flux than that exhibited in previous outbursts. Unfortunately, there is no strictly simultaneous coverage of this flare in any other band.

\begin{figure}[H]
\centering
\includegraphics[width=0.55\textwidth]{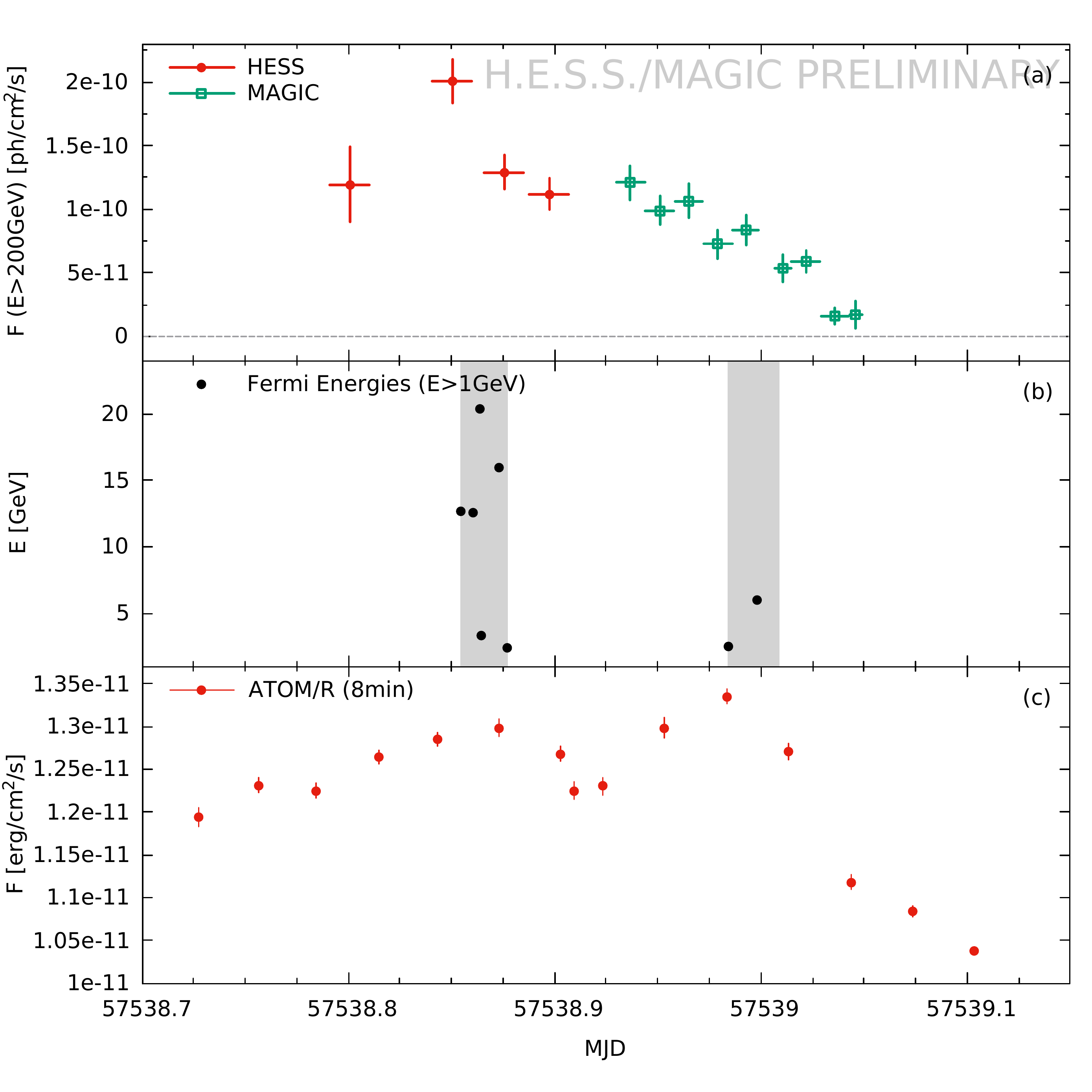}
\caption{(\textbf{a}) VHE lightcurve obtained with H.E.S.S. (red) and MAGIC (green) for $E>200\,$GeV in 28\,min and 20\,min bins, respectively, for the maximum night, May 30th, 2016 (MJD~$57538$). (\textbf{b})~Individual photons detected with \fermi\ with the gray bands indicating the visibility window of PKS~1510-089 for {\it Fermi}. (\textbf{c}) Optical R-band lightcurve obtained with ATOM in 8\,min bins. In all panels only statistical errors are shown.}
\label{fig:flarelc-57538}
\end{figure}

A detailed lightcurve of the maximum night is shown in Figure~\ref{fig:flarelc-57538}. The VHE flux exhibits a peak with a flux $\sim$80\% of the Crab above an energy of $200\,$GeV and a subsequent decay. From the peak to the minimum the flux fell by almost an order of magnitude. As the low flux in the HE band coupled with the small effective area of \fermi\ inhibits short-time binning, individual photons recorded with \fermi\ with energies $E>1\,$GeV are shown in the second panel. \fermi\ recorded photons with energies up to $E\sim 25\,$GeV during the H.E.S.S. observation window, but only 2 photons with energies $E>1\,$GeV in the MAGIC observation window. This is indicative of a softening of the spectrum at that time. The optical R-band flux recorded with ATOM exhibits a double-peaked structure, which is different than the VHE \g-ray lightcurve. The optical flux only changes by $\sim$30\%.
\unskip

\begin{figure}[H]
\centering
\includegraphics[width=0.50\textwidth]{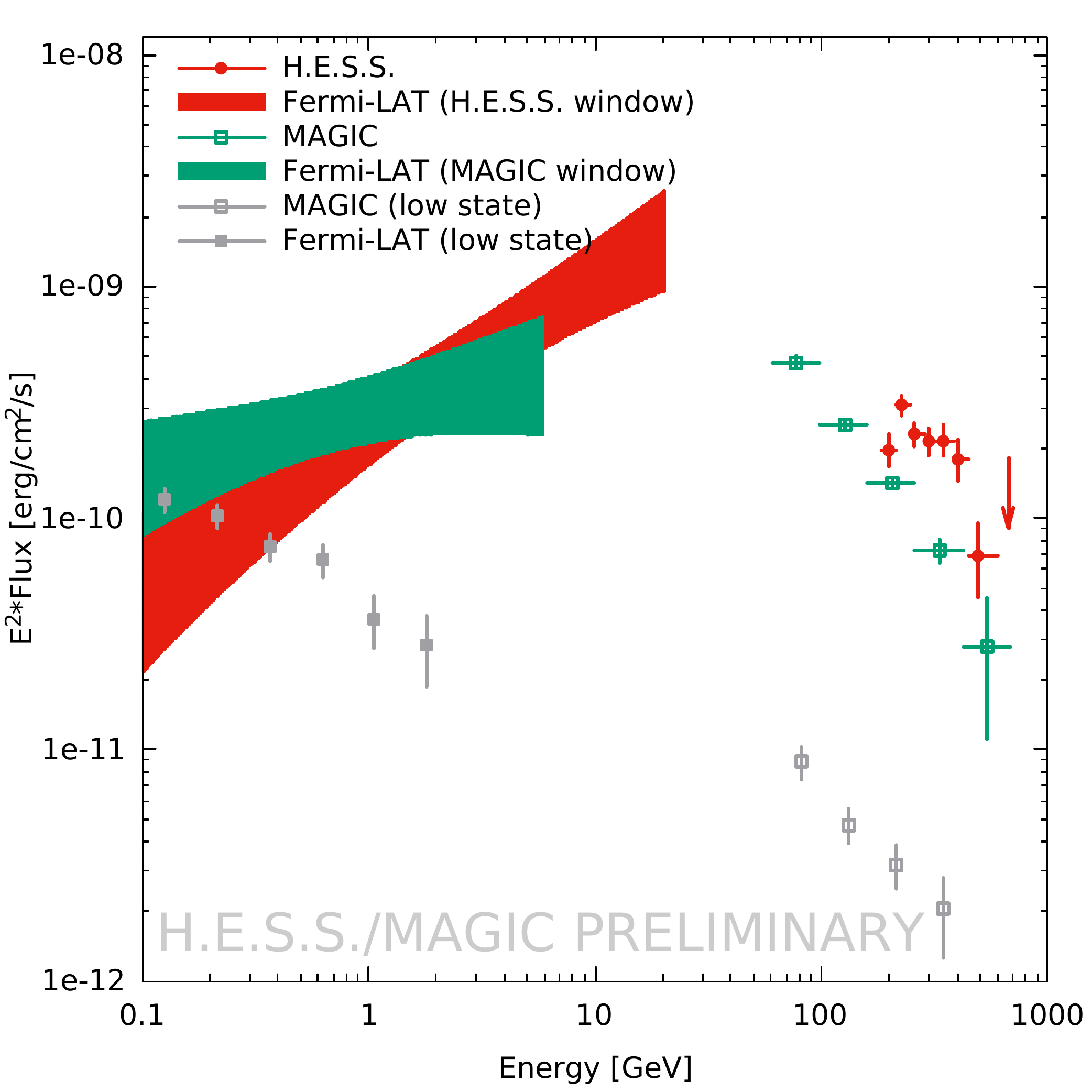}
\caption{$\gamma$-ray spectral energy distribution for the H.E.S.S. (red) and MAGIC (green) observation windows of the maximum night. The \fermi\ confidence regions have been derived for strictly simultaneous time windows with the IACT observations. The gray spectra are the low-state data sets of MAGIC and \fermi\ \cite{acc18}. The spectra have been corrected for EBL absorption using \cite{frv08}.}
\label{fig:gamma-spec}
\end{figure}

The $\gamma$-ray spectra of the maximum night are shown in Figure~\ref{fig:gamma-spec} along with the HE and VHE \g-ray low-state spectra \cite{acc18}. The VHE \g-ray spectra have been corrected for the absorption by the extragalactic background light (EBL) using the model of \cite{frv08}. The resulting deabsorbed spectra of the flare are compatible with power-laws with indices $\Gamma_{H.E.S.S.} = 2.9\pm 0.2_{\rm stat}$ and $\Gamma_{MAGIC}= 3.37\pm 0.09_{\rm stat}$, respectively. The HE \g-ray spectra in the two VHE observation windows are compatible within errors being $\Gamma_{LAT}=1.4\pm 0.2_{\rm stat}$ during the H.E.S.S. time frame and $\Gamma_{LAT} = 1.7\pm 0.2_{\rm stat}$ during the MAGIC time frame, respectively. The spectral breaks between the HE and the VHE $\gamma$-ray band are, therefore, $\Delta\Gamma = 1.5\pm 0.3_{\rm stat}$ during the H.E.S.S. time frame and $\Delta\Gamma = 1.7\pm 0.2_{\rm stat}$ during the MAGIC time frame. The comparison with the low-state data clearly shows the significant shift of the peak energy from $E\sim 100\,$MeV to $E\sim 30\,$GeV during this flare.

\begin{figure}[H]
\centering
\includegraphics[width=0.6\textwidth]{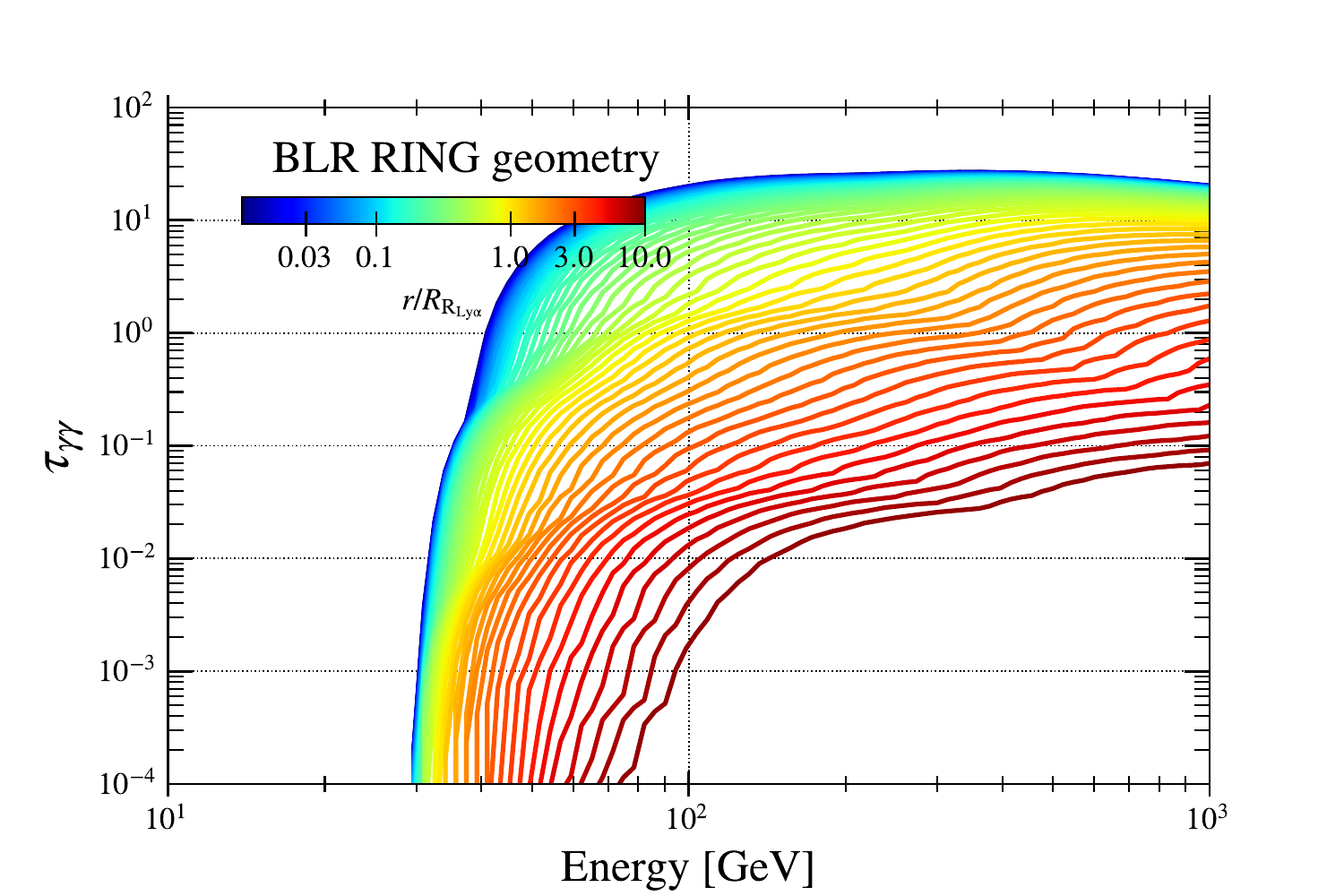}
\caption{The optical depth $\tau_{\gamma\gamma}$ as a function of $\gamma$-ray energy in the incident photon field of the BLR. The colors refer to different distances $r$ along the jet with respect to the distance $R_{{\rm Ly}_{\alpha}}\sim 8\times 10^{16}\,$cm of the Ly$_{\alpha}$ line.}
\label{fig:absorption}
\end{figure}

The spectral break can have several causes. The underlying particle distribution could exhibit a break due to the interplay of acceleration and cooling. The break could also be a sign of the Klein-Nishina reduction of the inverse-Compton cross-section at high energies. That the break remains roughly constant between the respective observation time windows while the spectra get softer, could~be an indication of the Klein-Nishina reduction. A third possibility is that the spectral softening results from the absorption of VHE photons by external soft photon fields, such as those from the BLR. The~resulting optical depth for different VHE emission region distances from the black hole is shown in Figure~\ref{fig:absorption}. Obviously, the distance of the emission region from the black hole has a strong influence on the optical depth $\tau_{\gamma\gamma}$.

One can conservatively estimate an upper limit on the degree of absorption by assuming that the spectrum detected by \fermi\ represents also the intrinsic spectrum in the VHE domain. The~degree of absorption $\tau$ can then be derived by
\begin{eqnarray}
 \tau = \ln{\frac{F_{extra}}{F_{obs}}} \label{eq:tau},
\end{eqnarray}
where $F_{extra}$ is the extrapolated flux, and $F_{obs}$ is the observed flux.

Without going into details \cite{zea16}, the calculation for the H.E.S.S. data set give a maximum value of $\tau$ for the highest energies of $\tau=5.4\pm0.9_{\rm stat}$, while the MAGIC data gives a maximum value of $\tau=3.9\pm1.4_{\rm stat}$. These estimates agree within errors. Assuming the absorption is due to the BLR, the~absorption values can be translated into a minimum distance of the emission region from the black hole. The emission region could be located at roughly $2\times R_{{\rm Ly}_{\alpha}} \sim 1.6\times 10^{17}\,$cm$\,\sim 0.05\,$pc from the black hole (c.f. Figure~\ref{fig:absorption}). Assuming that the distance of the Ly$_{\alpha}$ line represents the radius of the BLR~\cite{finke16}, the flaring region would at the very least be located on the outer edge of the BLR. This~underlines the statement given above during the discussion of the monitoring data: the jet must be able to produce VHE \g-rays on distances on the order of a significant fraction of a parsec from the black hole.

\section{Summary \& Conclusions}
FSRQs are by now established VHE \g-ray emitters. However, whether they are able to produce the VHE emission at all times or only during short bright flares has been an open question. This~has led to the establishment of monitoring programs by the IACT experiments H.E.S.S. and MAGIC on the FSRQ PKS~1510-089, which is one of the closest of this type of blazars (redshift $\zred=0.361$). These~programs are supplemented with multiwavelength data in the HE \g-ray, X-ray and optical regime. The first important result obtained by MAGIC is that PKS~1510-089 can be detected at VHE \g-rays during low states in the HE band \cite{acc18}. Similarly, variability has also been established through MAGIC observations \cite{aMea17}.

The H.E.S.S. monitoring described in detail here adds important features to these earlier results. Strong variability is detected in the VHE \g-ray domain, while the observations also hint to a persistent flux at other times. Interestingly, comparison of the VHE \g-ray lightcurve with other energy bands does not reveal any obvious correlation. This~exemplifies the need for deep monitoring programs across the entire multiwavelength spectrum. Otherwise important effects---such as better sampling of correlation functions, variability time scales, etc.---might be missed for the interpretation of certain~events.

The latter statement is further emphasized by the detection of an unprecedented VHE \g-ray flare in 2016 with H.E.S.S. that was followed up with MAGIC, as well. It was more than 10 times brighter than any flux seen at VHE \g-rays before (with a peak flux of $80$\% of the Crab) and lasted only 2 days. It was accompanied by a significant hardening of the HE \g-ray spectrum as observed with \fermi, while the HE fluxes remained rather low compared to other flares. The optical R-band observations with ATOM revealed a mild counterpart that was also much dimmer than previous flares, but exhibited a different flux evolution compared to the VHE band. Unfortunately, no other simultaneous data is available that could further constrain the spectrum.

All these observations reveal that the jet of PKS~1510-089 is able to accelerate particles to high energies to produce VHE \g-rays at all times. It also implies that these emission regions are probably located beyond the BLR, as otherwise the VHE emission should be strongly absorbed.
This has been shown here specifically for the 2016 VHE flare. A simple estimate of the maximum absorption allowed for by the data results in a lower limit on the black hole distance, which indicates a location of the emission region on the edge of or beyond the BLR.

In conclusion, deep and, preferably, unbiased monitoring programs on FSRQs and blazars in general are important to reveal the general behavior of the sources, as well as to uncover new and unexpected features.


\vspace{6pt}



\authorcontributions{Conceptualization, M.Z.; Data curation, E.L., M.M., J.S., S.W. and A.W.; Formal analysis, F.J., H.P., D.S. and A.W.; Investigation, M.Z., M.M. and J.S.; Methodology, H.P. and S.W.; Project administration, E.L.; Software, F.J. and M.M.; Validation, D.D.P., J.S., T.T. and A.W.; Writing---original draft, M.Z.
}

\funding{M.~Z. acknowledges funding by the German Ministry for Education and Research (BMBF) through grant 05A17PC3.
}

\acknowledgments{The support of the Namibian authorities and of the University of Namibia in facilitating the construction and operation of H.E.S.S. is gratefully acknowledged, as is the support by the German Ministry for Education and Research (BMBF), the Max Planck Society, the German Research Foundation (DFG), the~Alexander von Humboldt Foundation, the Deutsche Forschungsgemeinschaft, the~French Ministry for Research, the~CNRS-IN2P3 and the Astroparticle Interdisciplinary Programme of the~CNRS, the~U.K. Science and Technology Facilities Council (STFC), the IPNP of the Charles University, the~Czech Science Foundation, the~Polish National Science Centre, the South African Department of Science and Technology and National Research Foundation, the University of Namibia, the National Commission on Research, Science \& Technology of Namibia (NCRST), the Innsbruck University, the~Austrian Science Fund (FWF), and the Austrian Federal Ministry for Science, Research and Economy, the University of Adelaide and the Australian Research Council, the Japan Society for the Promotion of Science and by the University of Amsterdam.
We appreciate the excellent work of the technical support staff in Berlin, Durham, Hamburg, Heidelberg, Palaiseau, Paris, Saclay, and in Namibia in the construction and operation of the equipment. This work benefited from services provided by the H.E.S.S. Virtual Organisation, supported by the national resource providers of the EGI Federation.

MAGIC would like to thank the Instituto de Astrof\'{\i}sica de Canarias
for the excellent working conditions at the Observatorio del Roque de
los Muchachos in La Palma. The financial support of the German BMBF
and MPG, the Italian INFN and INAF, the Swiss National Fund SNF, the
ERDF under the Spanish MINECO (FPA2015-69818-P, FPA2012-36668,
FPA2015-68378-P, FPA2015-69210-C6-2-R, FPA2015-69210-C6-4-R,
FPA2015-69210-C6-6-R, AYA2015-71042-P, AYA2016-76012-C3-1-P,
ESP2015-71662-C2-2-P, CSD2009-00064), and the Japanese JSPS and MEXT
is gratefully acknowledged. This work was also supported by the
Spanish Centro de Excelencia ``Severo Ochoa'' SEV-2012-0234 and
SEV-2015-0548, and Unidad de Excelencia ``Mar\'{\i}a de Maeztu''
MDM-2014-0369, by the Croatian Science Foundation (HrZZ) Project
09/176 and the University of Rijeka Project 13.12.1.3.02, by the DFG
Collaborative Research Centers SFB823/C4 and SFB876/C3, and by the
Polish MNiSzW grant 2016/22/M/ST9/00382.
}

\conflictsofinterest{The author declares no conflict of interest.
}

%

\appendixtitles{no} 
\appendixsections{multiple} 
%
%

\reftitle{References}





\end{document}